\documentclass[10pt,letterpaper]{article}
\usepackage{opex3}

\begin{document}

\title{Directional free-space coupling from photonic crystal waveguides}

\author{Cheng-Chia Tsai$^{1,*}$, Jacob Mower$^{1}$, and Dirk Englund$^{1,2}$}

\address{$^1$Department of Electrical Engineering, Columbia University, New York, NY 10027, USA \\$^2$Department of Applied Physics and Applied Mathematics, Columbia University, New York, NY 10027, USA}

\email{*ct2443@columbia.edu} 



\begin{abstract}
We present a general approach for coupling a specific mode in a planar photonic crystal (PC) waveguide to a desired free-space mode. We apply this approach to a W1 PC waveguide by introducing small index perturbations to selectively couple a particular transverse mode to an approximately Gaussian, slowly diverging free space mode. This ``perturbative photonic crystal waveguide coupler'' (PPCWC) enables efficient interconversion between selectable propagating photonic crystal and free space modes with minimal design perturbations. 
\end{abstract}



\section{Introduction}
Photonic crystals (PCs) exhibit a photonic bandgap that enables localization and manipulation of light at the wavelength-scale. \cite{1}. In two-dimensional (2D) photonic crystals, an important element is the photonic crystal waveguide, which enables compact optical delays \cite{2,14}, slow light and enhanced nonlinear processes \cite{4,7}, and efficient connections between various on-chip devices \cite{5,3,6}. 

Various techniques exist for coupling to photonic crystal waveguides, including chip-to-fiber butt-coupling \cite{8}, directional coupling from a tapered fiber \cite{9,10}, and vertical grating couplers including slanted \cite{11} or focusing lens gratings \cite{12,13}. Although these gratings can reach high efficiency, they constrain the PC device design as they disrupt the PC lattice and typically affect all waveguides modes. Here, we introduce a general approach for introducing highly non-invasive grating couplers that rely on small perturbations to existing device designs. For instance, in a W1 photonic crystal waveguide in a holey PC membrane, the grating consists only of small dielectric index perturbations to a set of holes near the waveguide region. We call this coupling device a Perturbative Photonic Crystal Waveguide Coupler (PPCWC). Although this approach is general, we focus here on the situation where we introduce extra holes adjacent to the existing air holes. This restricted set of perturbations eliminates fabrication challenges that may be associated with non-circular or very small, isolated perturbations.  

\begin{figure}[b]
\centering
\includegraphics[width=12cm]{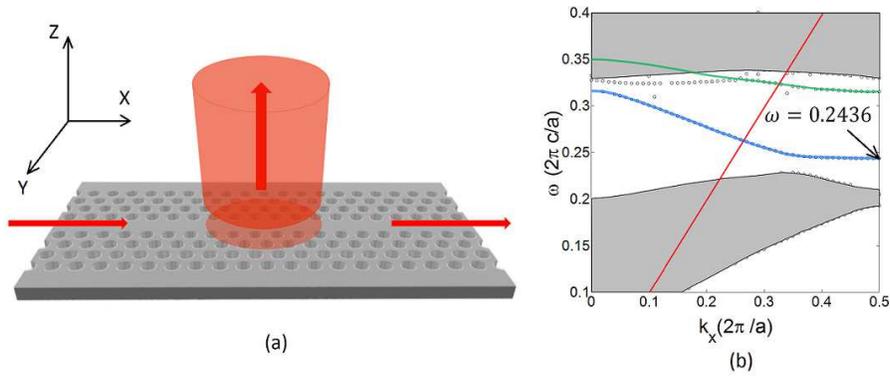}
\caption{(a) Schematic showing the coupling system. (b) The dispersion relation of a W1 PC waveguide. The fundamental mode (the blue line) and a higher mode (the green line), having frequencies $\omega=0.2436$ and $0.3151 (2\pi c/a)$, respectively, at wave-vector $k_x=\pi/a$, are labeled in the figure.}
\label{fig:2}
\end{figure}

Similar directional coupling from photonic crystal cavities by introducing small perturbation has been explored in previous studies \cite{19,15,18}. The index perturbation is derived here using coupled mode theory, under the constraint that it can be easily integrated into the photonic crystal fabrication. This approach enables a straightforward design of perturbations that causes selective coupling between a desired PC waveguide mode and a desired free-space mode (chosen here to be a Gaussian). Figure \ref{fig:2}(a) shows the coupling scheme. When light propagates along the waveguide (the red arrows) and passes through the coupler, the periodic index perturbations act as a grating and light scatters out of the slab. We use three-dimensional finite difference time domain (FDTD) simulations to calculate the waveguide modes for a given index perturbation and to find the efficiency of the resulting grating coupler. We find that even very small perturbations allow a high coupling efficiency between the free-space radiative mode and a given waveguide mode, while causing only very small changes to the propagation of other waveguide modes. As demonstrated in the end of Section 2, the directional scattering from this coupler is controllable by detuning the feed frequency or modulating the perturbation size. In Section 3, we discuss the mode selectivity of our device by investigating the problem in $k$-space, and in Section 4 we analyze the transmittance and the modification of the shape and the directionality of the scattered beam. 

\section{Coupled mode theory}
We consider a guided wave in a W1 photonic crystal line defect waveguide and the light scattered at the grating coupler. We derive the coupled mode equations from wave equation
\begin{equation}
\nabla \times \nabla \times \vec{E}\ = \frac{\varepsilon(\vec{r})}{c^{2}}\frac{\partial ^{2} \vec{E}}{\partial t^{2}}, 
\end{equation}
where $\varepsilon(\vec{r})$ is the relative dielectric constant and \textit{c} is the speed of light in vacuum. We define $\varepsilon_w$ as the relative dielectric constant for the waveguide, which is assumed to be periodic along the propagating direction. The solution of Eq. (1) with $\varepsilon=\varepsilon_w$ is in form of
\begin{equation}
\vec{E_w}=\vec{B_w}(\vec{r})e^{i(\omega(\vec{k})t-\vec{k}\cdot\vec{x})}.
\end{equation}

$\vec{B_w}$ are Bloch states having periodicities as the multiples of the periodicity of $\varepsilon_w$, where $\vec{k}$ represents the crystal momentum and $\hat{x}$ is the propagating direction of the waveguide. We desire a scattered beam propagating in free space denoted by $\vec{A}(\vec{r})$. $\vec{A}$ is also a solution of Eq. (1) with the relative dielectric constant $\varepsilon=\varepsilon_G$. We assume the field to be a superposition of the waveguide and free space modes: 
\begin{equation}
\vec{E}=a(t)\vec{A}(\vec{r})e^{i\omega_G t}+\sum_{\vec{k}}{\vec{B_{\vec{k}}}(\vec{r})e^{i\omega(\vec{k}) t}[b(\vec{k},t)e^{-i\vec{k}\cdot\vec{x}}+c(\vec{k},t)e^{i\vec{k}\cdot\vec{x}}]}.
\end{equation}

Here, $a(t)$ is the slowly varying component of the Gaussian beam, and $b(\vec{k},t)$ and $c(\vec{k},t)$ are the slowly varying components of the forward and backward propagating Bloch modes for the waveguide, respectively. Substituting Eq. (3) into Eq. (1), we can obtain the full coupled mode equations describing the interaction of the coefficients $a(t)$, $b(\vec{k},t)$, and $c(\vec{k},t)$ through $\Delta \varepsilon$. In particular, the coupling rate between $a(t)$ and $b(\vec{k},t)$ can be evaluated \cite{3,17} to be
\begin{equation}
\kappa_{ba}(\vec{k})= \frac{\int{dr\frac{\Delta\varepsilon_w(\vec{r})}{c^2}e^{-i\vec{k}\cdot\vec{x}}\vec{A^*}\cdot\vec{B_{\vec{k}}}}} {2\int{dr\frac{\varepsilon_t(\vec{r})}{c^2}|\vec{A}|^2}}
\end{equation}
where $\Delta\varepsilon_w=\varepsilon_t-\varepsilon_w$, representing the index perturbations introduced into the already existing structures. The subscript letters $w$ and $t$ denote the existing waveguide and the target structure. 

\begin{figure}[t]
\centering
\includegraphics[width=9cm]{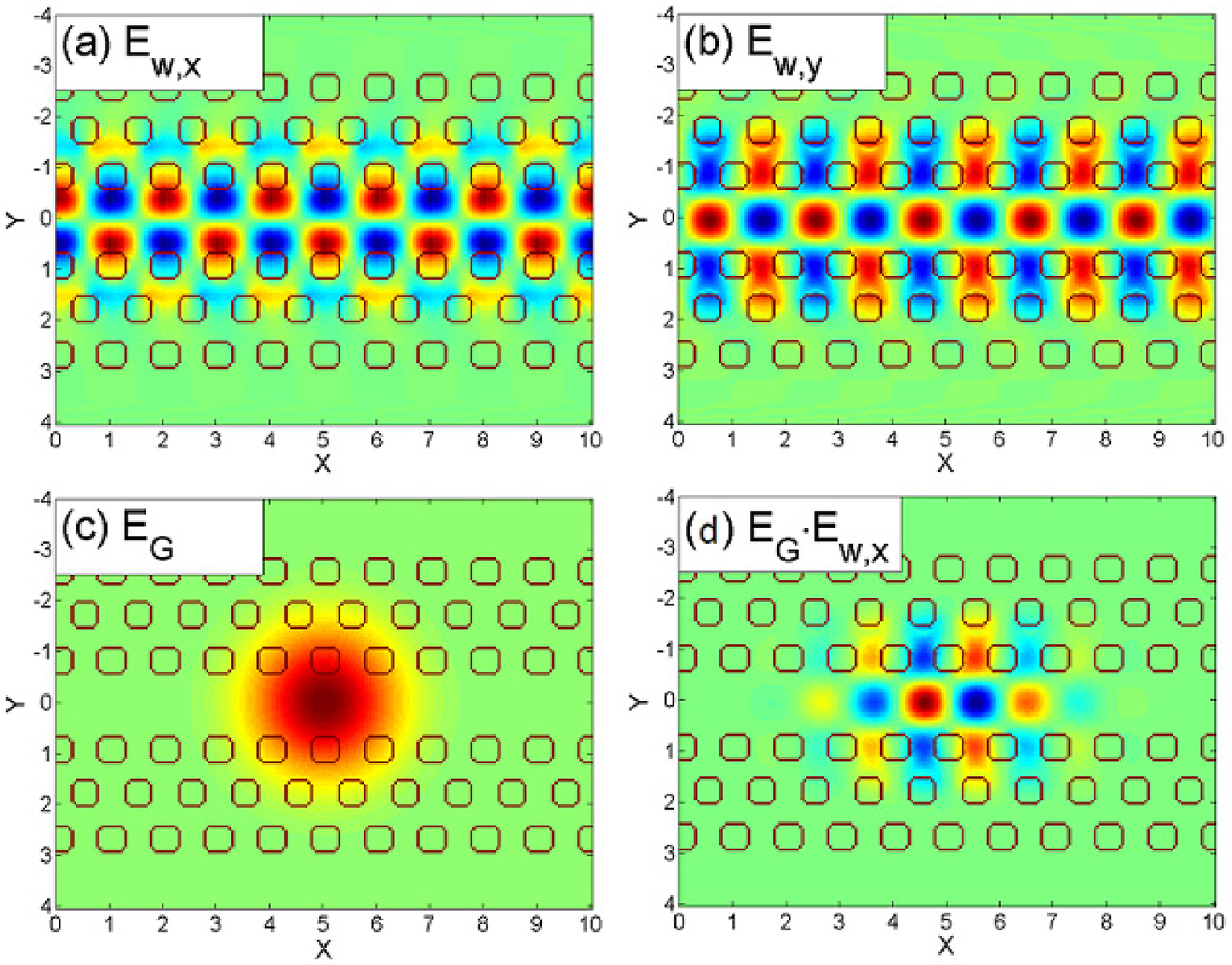}
\centering
\includegraphics[width=10cm]{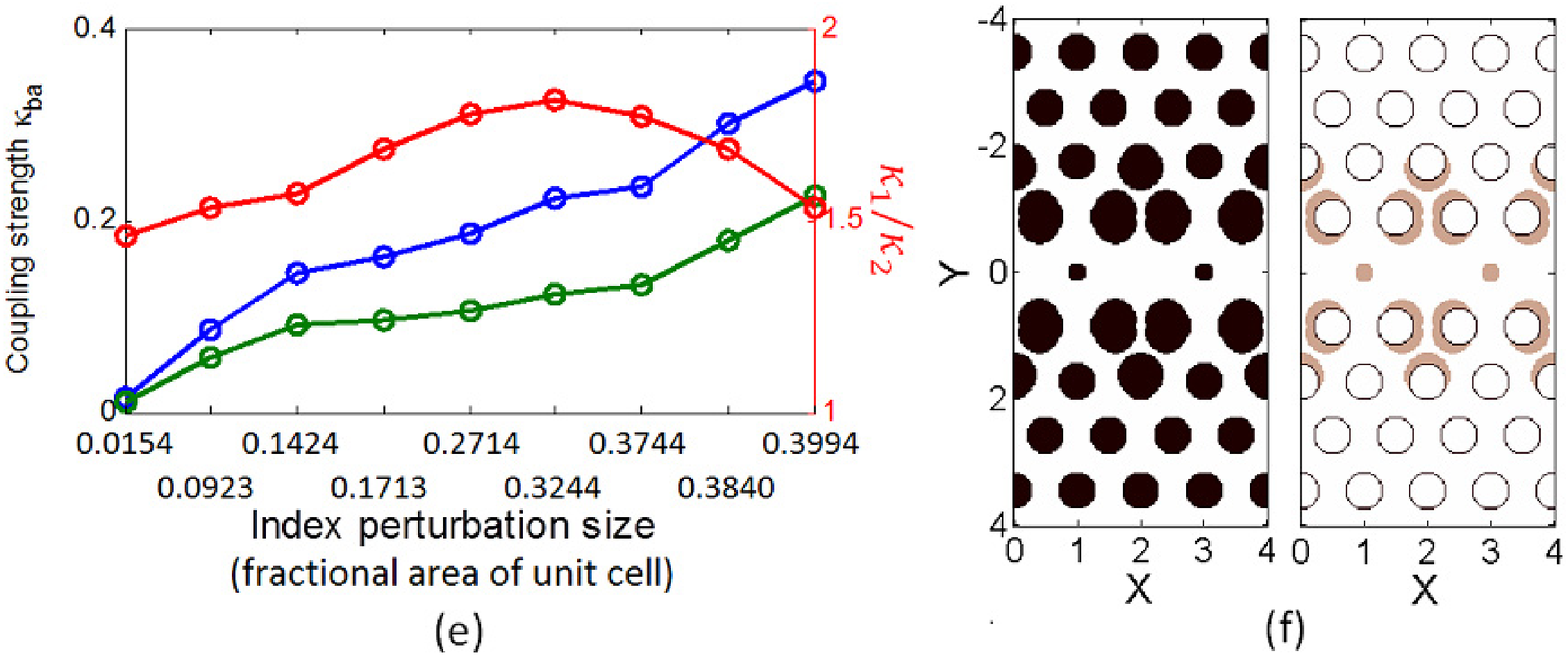}
\caption{FDTD Simulation of PC waveguide and calculated coupling strength of index perturbed vertical coupler. (a) (b) $\hat{x}$ and $\hat{y}$ components of electric field of the waveguide mode $\omega=0.2436$. (c) A Gaussian beam profile which we expect as the scattered light. (d) Dot product of the $\hat{y}$ component of the waveguide mode and the target Gaussian mode. (e) Calculated coupling strength $\kappa_{ba}$ (Eq. (4)) for target Gaussian modes with different divergences. (f) The resulting perturbed PC structure (the left panel) and the index perturbation distribution (the right panel) of the case with the perturbation size of 0.3244, in units of the fractional area of unit cell, in (e).}
\label{fig:1}
\end{figure}
From Eq. (4), the coupling constant is determined by $\vec{A^*}\cdot \vec{B_{\vec{k}}}$ and the perturbation term $\Delta\varepsilon_w$. Since the phase of A is nearly constant in the plane of the PC, whereas $\vec{B_{\vec{k}}}$ varies rapidly along the direction of propagation, the spatial integral of the dot product $\vec{A^*}\cdot \vec{B_{\vec{k}}}$ will vanish. For a non-vanishing coupling constant, the index perturbation $\Delta\varepsilon_w$ should have the same periodicity as $\vec{B_{\vec{k}}}$, assuming an interaction region that is large compared to the period $a$. 

Now we consider a realistic waveguide made by introducing a line defect in a hexagonal photonic crystal lattice with periodicity $a$, refractive index of the slab material $n=3.6$, slab thickness $h=0.7a$, and hole radius $r=0.3a$. The full coupling system is as shown in Fig. \ref{fig:2}(a). We used a 3D FDTD simulation to calculate the waveguide modes. Figure \ref{fig:2}(b) shows the dispersion relation of the W1 photonic crystal waveguide. A fundamental and higher order mode inside the photonic bandgap (PBG), wedged by the areas shaded in gray, are shown in blue and green, respectively, and are confined below the light line, represented by the red line. Figures \ref{fig:1}(a) and \ref{fig:1}(b) show the $\hat{x}$ and $\hat{y}$ electric field components of the fundamental mode, which we label as $\vec{B_{\vec{k}}}$, at wave-vector $k_x=\pi /a$, and frequency $\omega=0.2436$, in units of $2\pi c/a$, where c is the speed of light in free space. As shown previously, the index perturbation must have a period of $2\pi/\vec{k_x}=2a\hat{x}$ to scatter this $k$-state to the $\Gamma$ point in $k$-space, where the in-plane $k_x=0$. Note that because the $\hat{x}$ and $\hat{y}$ components of the electric field of the waveguide mode have different symmetries along the $\hat{y}$ direction, an index perturbation with even symmetry across $y=0$ will scatter only the $\hat{y}$-component, but not the $\hat{x}$-component of the field. Similarly, it is possible to scatter only the $\hat{x}$-component with a anti-symmetric perturbation distribution across the $y=0$ line. For the following analysis, we will choose to scatter the $\hat{y}$-component. Furthermore, we suppose that we want to scatter into a vertical mode that has a Gaussian profile in the in-plane directions, as shown in Fig. \ref{fig:1}(c). Figure \ref{fig:1}(d) shows the dot product $\vec{A^*}\cdot\vec{B_{\vec{k},y}}$, which appears in the integrand in Eq. (4). To find a perturbation that scatters light into a slowly diverging Gaussian beam, we calculate the coupling rate for beams with different divergence angles and compare the coupling constants among different index perturbation designs. We discuss in this paper, the perturbations are formed by enlarging three-hole clusters with periodicity $2a$ in the $\hat{x}$-direction. The blue and green lines in Fig. \ref{fig:1}(e) represent the coupling constants $\kappa_1$ and $\kappa_2$ for Gaussian beams, focused in the plane of the photonic crystal, with waists of $2a$ and $a$, respectively, and the red line represents the ratio $\kappa_1 / \kappa_2$. The horizontal axis denotes the perturbation size in units of the area of the unit cell of the triangular lattice, and the vertical axes denote the coupling constant (left) and the ratio $\kappa_1/\kappa_2$ (right). We observe that a local maximum exists for $\kappa_1/\kappa_2$ when the index perturbation size is 0.3244, in units of the fractional area of unit cell, that is, the light scattered in this case diverges less than a well-defined Gaussian beam. The perturbed structure for this case is shown in Fig. \ref{fig:1}(f) and the corresponding coupling strength $\kappa_1$ and the ratio $\kappa_1/\kappa_2$ are 0.2228 and 1.815, respectively. As expected, the coupling rate into the Gaussian beam of smaller waist is lower than for the wider, less rapidly diverging beam. The brown shaded area in the right panel of Fig. \ref{fig:1}(f) shows the index perturbation. 

\begin{figure}[t]
\centering
\includegraphics[width=9cm]{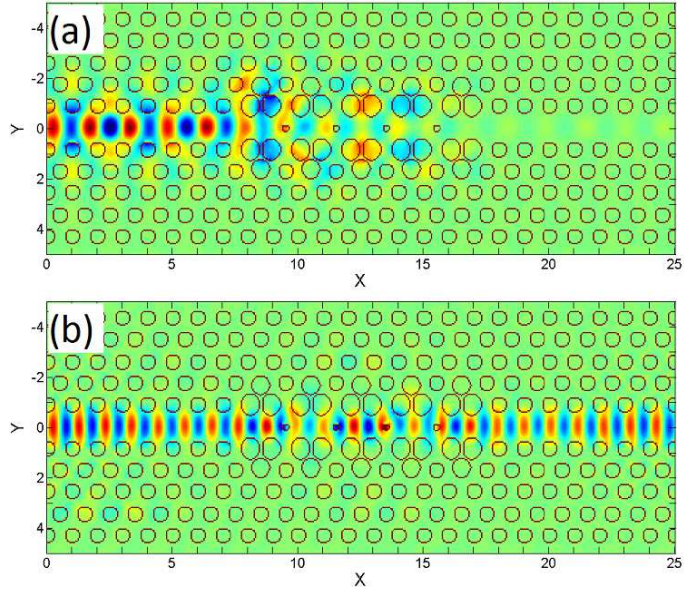}
\caption{Structures and simulation results of a line defect PC waveguides with and index perturbation introduced. Figures shows the distributions of the $\hat{y}$ component of the electric field of the waveguide modes at (a) $\omega=0.2436$ and (b) $\omega=0.3078$, respectively.}
\label{fig:3}
\end{figure}
\begin{figure}[t]
\centering
\includegraphics[width=8cm]{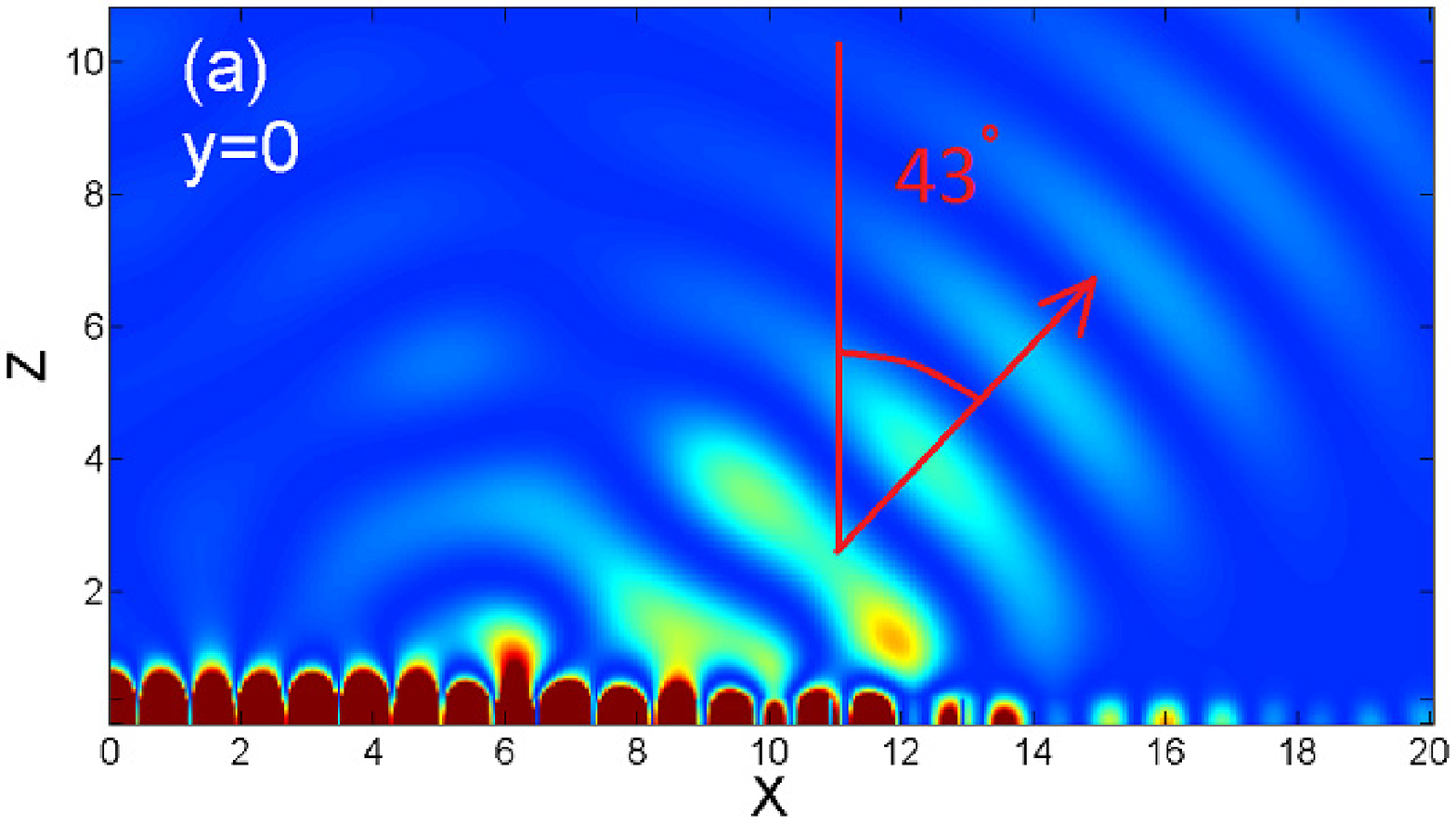}
\centering
\includegraphics[width=11cm]{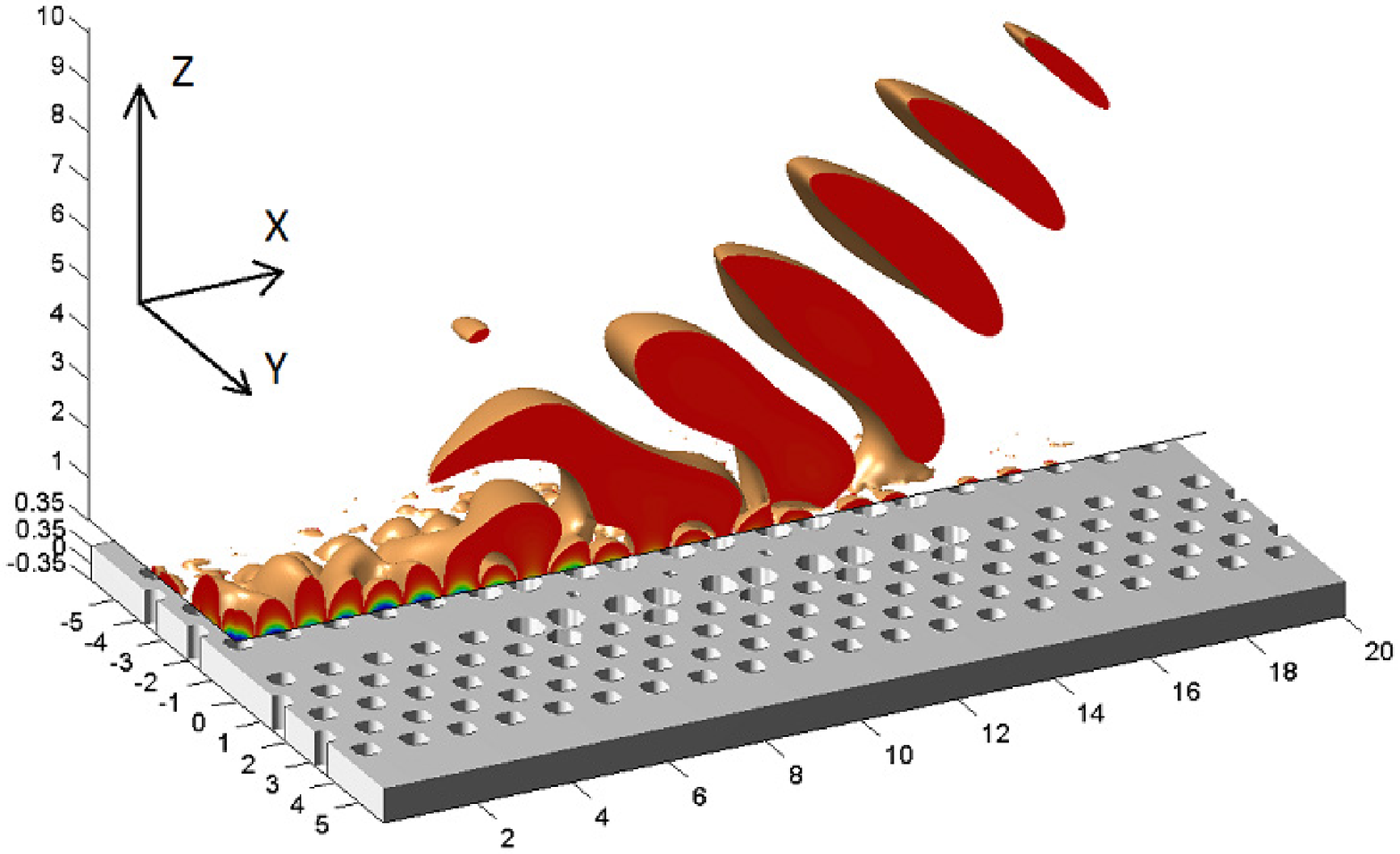}
\caption{(a) The intensity distributions of the $\hat{y}$ component of the electric field, $E_y$, of the scattering from a 5-period long coupler, of which the perturbation pattern is as shown in \ref{fig:1}(f), viewed in a vertical plane passing through the middle of the waveguide, where $z=0$ is the center of the PC slab. The waveguide is excited at $\omega=0.2436$. (b) The three dimensional view of the density isosurface of the $\hat{y}$ component of the electric field, $E_y$. The structure of the PC waveguide with a coupler is shown at the bottom.}
\label{fig:4}
\end{figure}
We check our theoretical predictions using another 3D FDTD simulation of the coupling efficiency of a grating consisting of the index perturbation as shown in Fig. \ref{fig:1}(f). The fundamental waveguide mode at frequency $\omega=0.2436$ is excited by a Gaussian beam source at one end of the PC waveguide. The $\hat{y}$-component of the electric field in the plane of the PC is shown in Fig. \ref{fig:3}(a), which contains five periods of the index perturbation shown in Fig. \ref{fig:1}(f). The perturbation couples the incident waveguide mode upward, resulting in reduced waveguide transmission ($<0.5\%$) on the right side of the perturbation. In Fig. \ref{fig:3}(b), we consider the effect of this perturbation on a higher-frequency mode at $\omega=0.3078$, which also has even symmetry across the $y=0$ axis, but which has a periodicity of $a$ in the direction of propagation. Equation (4) predicts that if $\Delta\epsilon$ has a periodicity of $2a$, the coupling constant $\kappa_{ba}$ will vanish in this case. This result is confirmed by a high in-plane transmission after the coupler ($>99\%$), as shown in Fig. \ref{fig:3}(b). We will discuss this mode selective property of the coupler in greater detail in Section 3.

\begin{figure}[t]
\centering
\includegraphics[width=13cm]{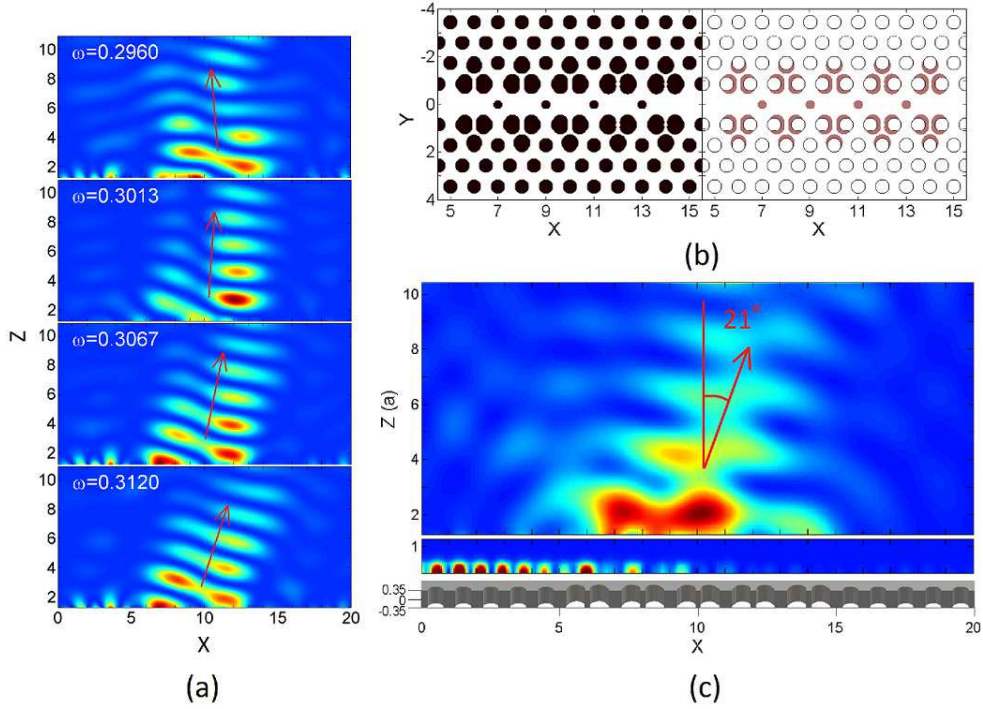}
\caption{The control of the scattering direction. (a) Intensities of the scattered fields at different feeding frequencies. (b) The structure of the a 5-period long coupler with increasing perturbation size. The brown shade area in the right panel labels the perturbation introduced. (c) The intensity distribution of the scattering field radiated from the coupler as shown in (b). The bottom panel is a $x-z$ cross-section through the center of the first row of holes.}
\label{fig:6}
\end{figure}
Figure \ref{fig:4} displays the energy density of the $\vec{y}$ component of the evanescent field and the upward component of the scattering field from the coupler. Figure \ref{fig:4}(a) is a snapshot viewed in the vertical cross-section through a plane passing through the middle of waveguides, and Fig. \ref{fig:4}(b) is the three dimensional view of the $E_y$ energy density isosurface. We observe a strong radiated field originating from the 5-period long perturbed region. 

As observed in Fig. \ref{fig:4}, the radiation exits at $43^\circ$ from the vertical. This direction depends on the interference of the fields scattered by discrete perturbations, as in a line grating. The interference is governed by relative amplitudes and phases of the fields at the positions of the perturbations. The scattered field can be approximated as a sum over all perturbations:
\begin{equation}
\vec{E}(\vec{r})=\sum_{i=1}^N{\vec{E_i}(\vec{r})e^{i\vec{k_{w}}\cdot(\vec{r_i}-\vec{r_1})-\alpha|\vec{r_i}-\vec{r_1}|}},
\end{equation}
where
\begin{equation}
\vec{E_i}(\vec{r})=\frac{\vec{u}(\vec{r}-\vec{r_i})}{|\vec{r}-\vec{r_i}|}e^{i\vec{k_0}\cdot(\vec{r}-\vec{r_i})}
\end{equation}
represents the propagation of the radiation from a scattering point at position $\vec{r_i}$, $\vec{u}(\vec{r})$ denotes the spatial profile of the radiated field, $\vec{k_0}$ is the $k$-vector in free space, $\vec{k_{w}}=(\pi/a+\Delta k_{w,x})\hat{x}$ is the crystal momentum of the perturbed PC waveguide where $\Delta k_{w,x}$ indicates a possible shift in the $k$-vector due to a slightly different dispersion in the perturbation region. $\vec{r_i}$ is th position of the $i$-th scattering point, and $\vec{r_1}$ is the position of the start of the coupler. The field dependence in Eq. (5) includes the decrease of the field amplitude along the coupler, $\alpha$, and the phase difference of the waveguide mode between perturbations. Because the perturbations are $\vec{r_i}$ have a periodicity of $2a\hat{x}$, the phase difference is equivalent to $2a\Delta k_{w,x}$. Equation (5) therefore predicts a shifting of the diffraction angle with wavelength, as given by the waveguide's dispersion relation. We observe this dependence in numerical simulations. In Fig. \ref{fig:6}(a), the frequency of the incident waveguide mode is tuned from $\omega=0.2960$ to $\omega=0.3120$ as the scattering angle shifts from $-4^\circ$ to $20^\circ$.

The coupled mode theory discussed above assumes a weak perturbation to the waveguide mode. However, the mode intensity drops as the field propagates through the coupler. This decrease in mode intensity should be taken into account to improve the efficiency of the coupler. Figure \ref{fig:6}(b) shows the structure of a 5-period long coupler in which we now include an increasing perturbation size to compensate for the decreasing field intensity. Using Eq. (5,6) to estimate the field intensity profile along the waveguide, we simulate a $5$-period long coupler with perturbation sizes multiplied by factors of $0.8589$, $0.8973$, $0.9059$, $0.9731$, and $1$. Figure \ref{fig:6}(c) shows the resulting intensity distribution of the scattered field viewed in a vertical plane through the center of the waveguides. We now observe a more uniform scattering along the length of the perturbation coupler, which results in higher mode overlap with the vertical Gaussian beam. 

\begin{figure}[t]
\centering
\includegraphics[width=13cm]{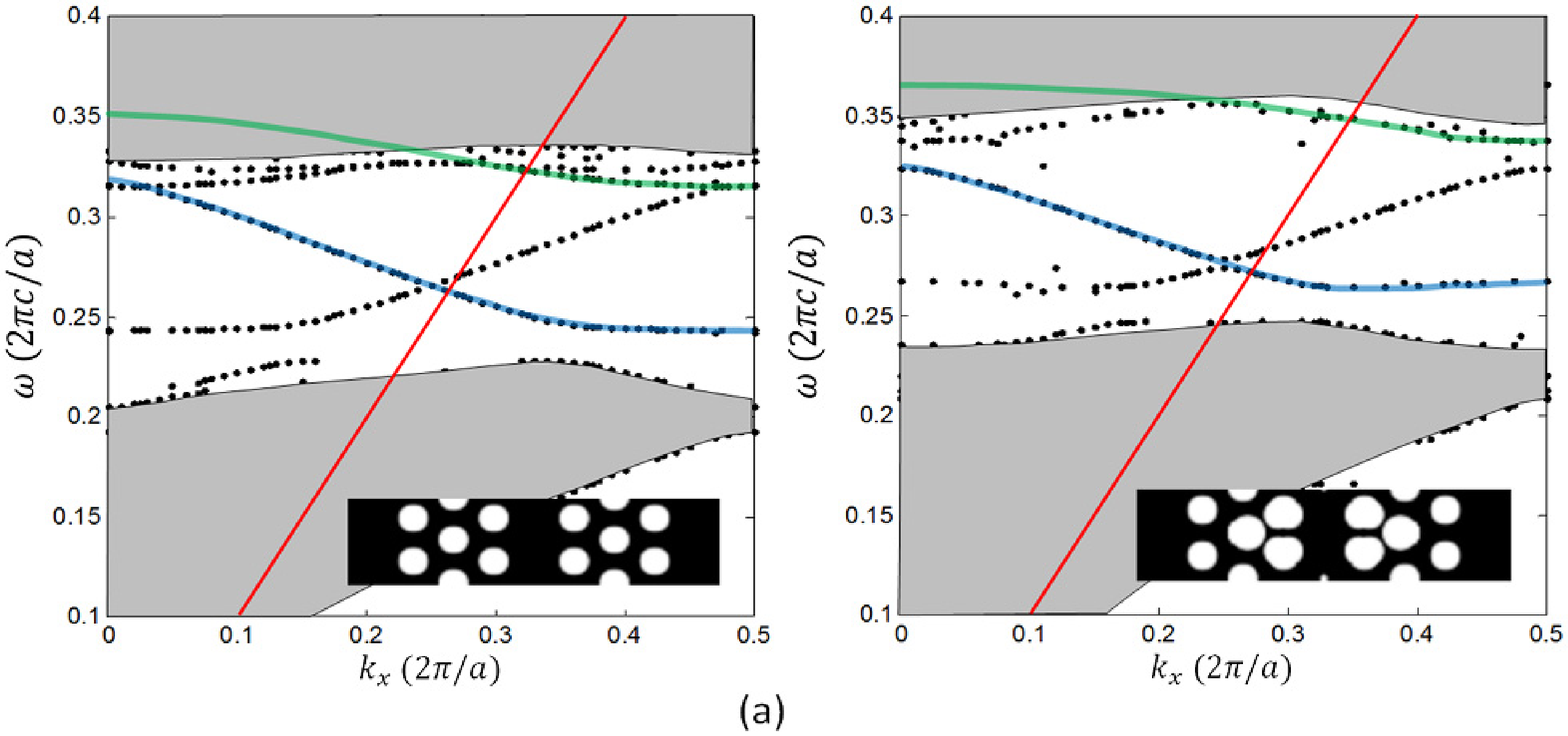}
\centering
\includegraphics[width=13cm]{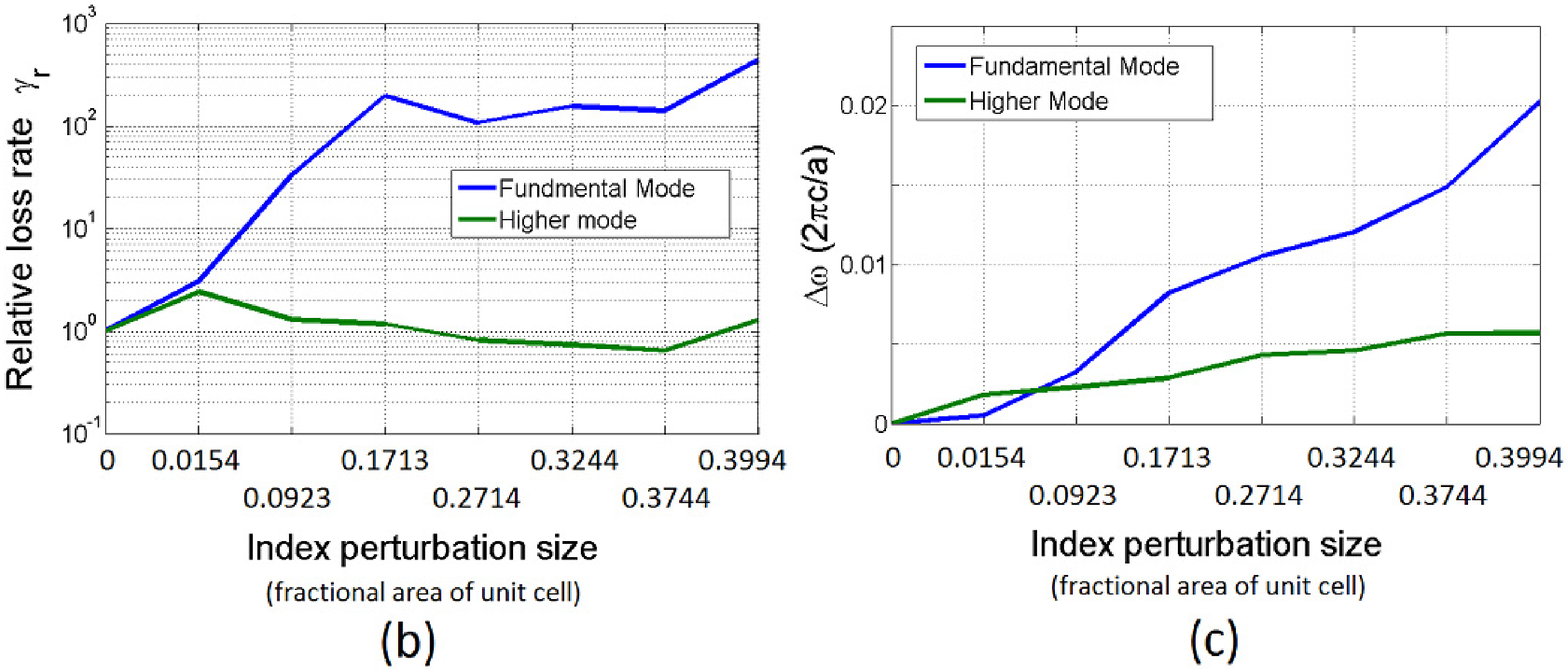}
\caption{(a) Dispersion relations of waveguides without (left) and with index perturbation (right). The insets show the unit cells of the periodic structure for each case. The perturbed structure is the same as shown in Fig. \ref {fig:1}(f). (b) Comparison of the changes in the relative decay rate of the fundamental mode (the blue line) and a higher mode (the green line). (c) Comparison of the changes in mode frequency of the fundamental mode (the blue line) and a higher mode (the green line).}
\label{fig:5}
\end{figure}

\section{Mode selectivity}
It is clear from the coupled mode picture that the out-coupling rate from the waveguide depends on the in-plane periodicity and the symmetry of the waveguide mode; for example, as was shown in Fig. \ref{fig:3}, the in-plane transmission after the coupler is suppressed for the fundamental mode, but not for the higher-order mode at $\omega=0.3078$. We now analyze this mode selectivity more closely in $k$-space. Figure \ref{fig:5}(a) compares the dispersion relations for the fundamental mode (the blue line) and a higher mode (the green line) inside the photonic band gap, having frequencies of $\omega_A=0.2436 (2\pi c/a)$ and $\omega_B=0.3151$ at $k_x=\pi/a$. For the structure with index perturbations (the right panel), the dispersion is raised slightly because of the decrease in the effective refractive index.  The periodicity of $2a$ scatters modes near $k_x=\pi/a$ by a lattice vector of $\Delta k_x=\pi/a$ to the $\Gamma$ point where $k_x=0$. Since these components are above the light line, they leak out of the PC plane and result in higher loss rate of the waveguide mode. 

In Fig. \ref{fig:5}(b), the blue line shows the loss rate, $\gamma_r$, normalized by the loss rate of the fundamental waveguide mode in the unperturbed structure, through different perturbation patterns. The green line shows $\gamma_r$ for the higher-order waveguide mode at $\omega=0.3151$. As the perturbation size increases, the loss rate of the fundamental mode increases by over two orders of magnitude, while the loss rate of the higher-mode remains roughly constant. This result demonstrates a high degree of selectivity of the perturbation coupler between different kinds of waveguide modes. Figure \ref{fig:5}(c) shows the frequency increase of the fundamental and higher order mode (the blue and green lines).

\begin{figure}[b]
\centering
\includegraphics[width=11cm]{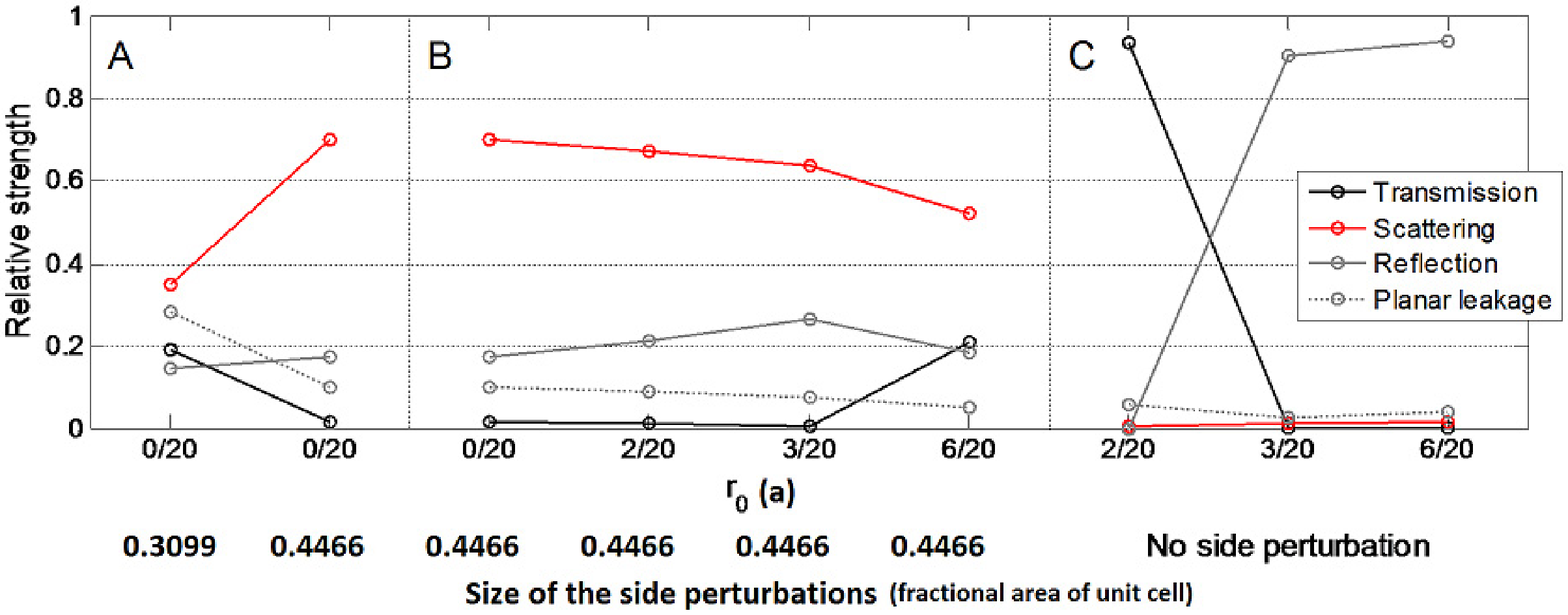}
\centering
\caption{The calculated relative strength of transmission (black solid line), scattering (red solid line), reflection (gray solid line), and planar leakage (gray dotted line) of the 5-period long couplers with different index perturbation distributions introduced. The horizontal axis denotes the perturbation size. $r_0$ is the radius of holes at the center of the waveguide, and the perturbation size introduced on the side is in units of the fractional area of unit cell.}
\label{fig:8}
\end{figure}

\section{Transmittance and beam shape modification}
The efficiency of the coupler is estimated by calculating the transmittance from the waveguide mode to a target free space Gaussian spatial mode, which includes both upward and downward components. We approximate the efficiency by   
\begin{equation}
T_c=S \times \frac{|{\langle E_{scatt}\cdot E_{G}\rangle}|^2}{\langle E_{scatt}^2\rangle\langle E_{G}^2\rangle},
\end{equation}
where $S$ is the transmission out of the PC plane and the fraction quantifies the mode overlap of the upward(or downward)-traveling component of the radiated field with the target Gaussian beam profile. The scattering rate of the couplers with different index perturbations are shown in Fig. \ref{fig:8} (the red solid line). The perturbations not only scatter the incident waveguide mode into the Gaussian mode, but also induce backward reflection (the gray solid line) and other in-plane leakage (the gray dotted line). The maximal scattering rate is reached when $r_0 = 0$ and the perturbation size in the sides is 0.4466, in units of the fractional area of unit cell. As shown in Fig. \ref{fig:8}, the scattering rate doubles when the perturbation size increases about $40\%$ (the region A), but slightly decreases as the radius of the central holes increases (the region B). If a perturbation is introduced only at the center of the waveguide (the region C in Fig. \ref{fig:8}), then the incident light is mostly reflected and only weakly scattered out of plane. 

When coupling to free space optics, and especially for single mode fiber coupling, the scattered field should have a large overlap with a Gaussian mode. In the case with the maximal scattering rate of 0.6989, we calculate an overlap of $89\%$ with a Gaussian beam focused in the plane of the PC. This indicates that the coupler allows for both strong scattering and a highly Gaussian-like profile. The resulting transmittance through the coupler is $T_c=0.6989\times 89\%=62\%$.

\begin{figure}[t]
\centering
\includegraphics[width=12cm]{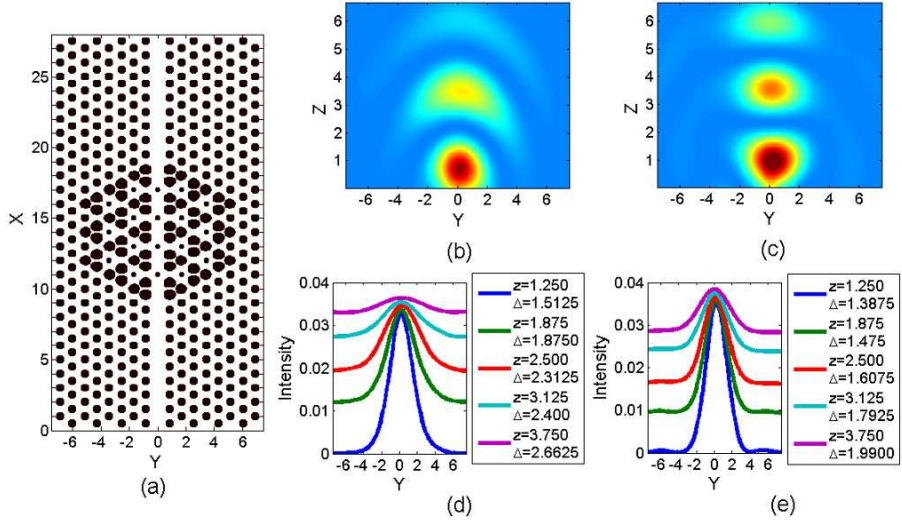}
\caption{(a) The structure of a coupler with additional index perturbation introduction to the 6th row from the waveguide. (b), (d) The intensity distribution viewed in a transverse plane through the center of the coupler shown in Fig. \ref{fig:3}. (c), (e) The intensity distribution of the scattering from a coupler as shown in (a). $\Delta$ is the beam width (FWHM) measured at different heights above the center of waveguide, in units of lattice constant, $a$.}
\label{fig:7}
\end{figure}

It is possible to decrease the beam divergence to allow for coupling to lower-NA lenses. By increasing the extent of the PPCWC in both $\hat{x}$ and $\hat{y}$ (transverse in-plane direction), we expand the coupling region and decrease the output beam divergence. This modified index perturbation is shown in Fig. \ref{fig:7}(a), again optimized for the mode at $\omega=0.2436$. Figures \ref{fig:7}(b) and \ref{fig:7}(c) show the intensity distributions of beams scattered from a small coupler (Fig. \ref{fig:3}) and a broadened one (Fig. \ref{fig:7}(a)), viewed in $y$-$z$ plane through the centers of the couplers. Figures \ref{fig:7}(d) and \ref{fig:7}(e) compare the divergence of the radiated beams in both cases by fitting a Gaussian beam profile to the data. We assume $2\theta=2W_0/z_0$, where $2\theta$ is the divergence angle of the beam and $W_0$ and $z_0$ are the beam waist and depth of the focus, respectively. In the case of the broadened coupler, the divergence is decreased by $20\%$ from that of the original smaller coupler. 

Unfortunately, the out-of-plane scattering from the PPCWC is symmetrically upward and downward, so both contribute to the total scattering rate. While large $T_c$ can be achieved by the methods described above, the overall efficiency of the PPCWC is so far limited to 50\%. The upward scattering efficiency is defined as the directionality, 
\begin{equation}
D=\frac{P_{U}}{P_{U}+P_D},
\end{equation}
where $P_U$ and $P_D$ are the powers radiated in the upward and downward directions, respectively. One method of increasing $D$ is to redirect the downward scattered field with a mirror. This may be achieved with a DBR structure, separated from the PC slab by some distance, $d$. When $d = 0.4667h$, the downward radiated light reflected by the mirror interferes with the upward radiated light constructively and a local maximum of $D=70\%$ occurs. When we move the mirror further from the slab $(d > 0.6h)$, there is minimal interference between two beams due to divergence of the beams. The upward scattering rate is close to the total scattering rate in this case, and we achieve $D=95\%$. More sophisticated methods may be used to increase directional scattering, such as only partial etching of the PPCWC. This has been successfully employed for photonic crystal cavities \cite{19,18} and channel waveguides \cite{20}.

\section{Conclusions}
We have designed and implemented a method for selectively coupling a planar photonic crystal waveguide mode to a target free space mode. The method proceeds as follows: (1) one chooses a waveguide mode to out-couple, (2) one chooses an index perturbation to couple the mode that has the appropriate periodicity and symmetry given by coupled mode theory, and finally, (3) one optimizes the perturbation for the desired out-coupled beam parameters. To demonstrate this approach, we focused on the problem of coupling a W1 waveguide mode to a slowly diverging Gaussian mode. We were able to ensure that the target waveguide mode was coupled with high rate, selectivity, and overlap with the target Gaussian mode.

This perturbative photonic crystal waveguide coupler (PPCWC) not only enables efficient scattering of selected waveguide modes, but also allows modifications in the shape and the direction of scattering. The angle of the scattered field can be shifted by detuning the mode frequency with respect to the lattice constant or introducing some modulation to the perturbation size. By breaking the vertical symmetry of the PC or using reflectors under the PC slab, it is possible to minimize downward propagation, directing most vertical scattering upward. The PPCWC enables a straightforward integration of vertical in/out- couplers which can be employed in a range of application from optical interconnects to photonic crystal light sources. Furthermore, the spectrally resolved light coupler could find application in light trapping for thin-film solar photovoltaics.

\section{Acknowledgment}
C.C.T. was partially supported by the Center for Re-Defining Photovoltaic Efficiency Through Molecule Scale Control, an Energy Frontier Research Center funded by the U.S. Department of Energy, Office of Science, Office of Basic Energy Sciences under Award Number DE-SC0001085. J.M. was supported by the DARPA Information in a Photon program, through grant W911NF-10-1-0416 from the Army Research Office.

\end{document}